\documentclass{article}
\usepackage{spconf}

\usepackage{amsmath, bbm}
\usepackage{amssymb}
\usepackage{amsfonts}
\usepackage{amsthm}
\usepackage{algorithm}
\usepackage{algorithmic}
\usepackage{nicefrac}
\usepackage{bm}
\usepackage{hyperref}
\usepackage{cite}
\usepackage[utf8]{inputenc} 
\usepackage[T1]{fontenc}
\usepackage{graphicx}
\usepackage[font=small]{caption}
\usepackage{tikz}
\usetikzlibrary{positioning}
\usepackage{pgfplots}
\usepackage{epstopdf}
\usepackage{subcaption}
\usepackage{multirow}

\newtheorem{Problem}{Problem}

\input{mysymbol.sty}

\title{Generative Adversarial Networks for \\Graph Data Imputation from Signed Observations}

\name{Amarlingam Madapu$^\star$, Santiago Segarra$^\dag$, Sundeep Prabhakar Chepuri$^\star$, Antonio G. Marques$^\ddag$
\thanks{This work was supported by the following grants:  Spanish MINECO TEC2016-75361-R and Instituto de Salud Carlos III DTS17/00158. E-mails: amarlingamm@iisc.ac.in, segarra@rice.edu, spchepuri@iisc.ac.in, antonio.garcia.marques@urjc.es }}
\address{$^\star$Indian Institute of Science, India.  $^\dag$Rice University, USA. $^\ddag$King Juan Carlos University, Spain}

\begin{document}
\ninept
\renewcommand{\baselinestretch}{0.98}
\maketitle
\begin{abstract}
We study the problem of missing data imputation for graph signals from signed one-bit quantized observations.
More precisely, we consider that the true graph data is drawn from a distribution of signals that are smooth or bandlimited on a known graph.
However, instead of observing these signals, we observe a signed version of them and only at a subset of the nodes on the graph.
Our goal is to estimate the true underlying graph signals from our observations.
To achieve this, we propose a generative adversarial network (GAN) where the key is to incorporate graph-aware losses in the associated minimax optimization problem.
We illustrate the benefits of the proposed method via numerical experiments on hand-written digits from the MNIST dataset.
\end{abstract}
\section{Introduction}\label{sec:intro}

Networks and network data are ubiquitous in contemporary applications. 
Every day, millions of devices are gathering unprecedented amounts of information with complex irregular structure, calling for new models and tools to understand and process the collected data. 
Representing the irregular structure using a graph and modeling the information as graph signals defined on the nodes of this graph has emerged as a rigorous and efficient alternative to deal with the intricacies of contemporary data. 
A broad range of graph signals exist, with meaningful examples including neural activity defined on the regions of a brain network, the spread of a rumor on a social network, and the delay experienced at each station of a subway network. 
In recent years, the processing of graph signals has attracted a lot of attention from the statistics, machine learning (ML), and signal processing (SP) communities, with relevant results including sampling and inpainting~\cite{chen_2015_sampling, marques_2016_sampling, anis_2016_sampling, chepuri2017graph,chamon_2018_sampling, romero_2017_reconstruction,ortiz2019sparse}, denoising~\cite{hammond_2011_wavelet, shafipour_2017_fourier,rey2019underpdecoder}, filtering~\cite{SandryMouraSPG_TSP14Freq,segarra2017filtering}, and deep graph convolutional architectures~\cite{defferrard2016convolutional,gama2019convolutionalgraphs, yang_2018_enhancing} for graph supported data.

While the progress achieved in recent years has been noteworthy, the existing results exhibit some limitations. 
Signals are typically continuous and most of the observation models are linear. 
More importantly, the majority of the schemes fail to account for the fact that modern datasets oftentimes have missing data or outliers that corrupt the observed (nodal) values. 
Indeed, data imputation mechanisms that leverage the graph structure and are able to deal with non-linearities are essential for the subsequent application of processing tools on the data, and are the subject of this work.

In particular, this paper puts forth a graph-regularized generative adversarial network (GAN) architecture for graph data imputation from quantized (signed) observations. 
More precisely, we consider a graph signal which is assumed to vary smoothly across a known graph. 
We further assume that the signal is not fully observed and that we only have access to the quantized version of the signal at a subset of the nodes (e.g., a few sensors sending quantized observations to a fusion center, or a few people in a social network responding to a yes/no question). 
Our approach is then to postulate a GAN architecture that exploits the structure of the underlying graph in learning how to impute (reconstruct) the unknown graph signal values. 

\noindent {\bf Related work and contributions.} 
Sampling and ulterior reconstruction is the most studied problem within the literature of SP and ML for graphs~\cite{chen_2015_sampling, marques_2016_sampling, anis_2016_sampling, chepuri2017graph,chamon_2018_sampling, romero_2017_reconstruction,ortiz2019sparse}. 
Typical approaches consider linear observation models and then assume that the sought signals are either smooth or bandlimited in the graph. 
However, those schemes face a number of challenges when the observations are quantized or follow a non-linear model. 
Deep learning architectures such as convolutional neural networks (CNNs) and GANs \cite{goodfellow2016deep} have been successfully used for the imputation and reconstruction of non-graph data \cite{yoon2018gain,khobahi2019deep}, especially in scenarios where the observation model is non-linear and a set of training samples exist. 
Equally important, recent works have looked at the generalization of GAN architectures to operate over graph signals, but for problems different from imputation \cite{kipf2016variational,de2018molgan}.
Motivated by the previous discussion, and inspired by the results in \cite{yoon2018gain}, this paper proposes a new GAN architecture able to work with (one-bit) quantized measurements and giving rise to reconstructed solutions that are either smooth or bandlimited in the supporting graph. 
Our contributions are twofold: 
i)~We incorporate the smoothness/bandlimited priors of graph SP methods into the GAN framework to construct a novel imputation method for graph signals; and 
ii)~We illustrate the performance of this method for a quantized signal model and compare it with state-of-the-art related approaches.



\section{Preliminaries and Problem Statement}
\label{sec:problem}

A weighted and undirected graph $\ccalG$ consists of a node set $\ccalN$ of known cardinality $N$, an edge set $\ccalE$ of unordered pairs of elements in $\ccalN$, and edge weights $A_{ij}\in\reals$ such that $A_{ij}=A_{ji}\neq 0$ for all $(i,j)\in\ccalE$. The edge weights $A_{ij}$ are collected as entries of the symmetric adjacency matrix $\bbA$ and the node degrees in the diagonal matrix $\bbD:=\diag(\bbA\bbone)$. These are used to define the (combinatorial) Laplacian matrix $\bbL:=\bbD-\bbA$.
	
The main focus of this paper is not on processing graphs, but signals defined on the nodes of a graph. Formally, let $\bbx=[x_1,...,x_N]^T \in\mbR^N$ be a graph signal in which the $i$th element $x_i$ denotes the signal value at node $i$ of $\ccalG$ with Laplacian $\bbL$. The assumption in graph SP is that the properties of the signal $\bbx$ depend on the supporting graph. A simple but effective alternative to achieve this is to define a smoothness metric that quantifies how well the signal matches the graph. The most widely used is the so-called quadratic total variation based on the Laplacian matrix, which is defined as
\begin{equation}\label{E:smoothness_laplacian_quad}
\mathrm{TV}_\ccalG^{\ell_2}(\bbx):=\bbx^T\bbL\bbx=\sum_{(i,j) \in \ccalE} A_{ij} (x_i - x_j)^2.
\end{equation}
Clearly, $\mathrm{TV}_\ccalG^{\ell_2}(\bbx)$ penalizes signals $\bbx$ for which neighboring nodes have very different values and, for constant signals, one has that $\mathrm{TV}_\ccalG^{\ell_2}(\bbone)=0$. 
Other meaningful definitions for the total variation include $\mathrm{TV}_\ccalG^{\ell_0}(\bbx):=\sum_{(i,j) \in \ccalE} A_{ij} \mathbbm{1}\{x_i \neq x_j \}$, with $\mathbbm{1}\{\cdot\}$ denoting the indicator function, which promotes piece-wise constant signals among densely connected communities and $\mathrm{TV}_\ccalG^{\ell_1}(\bbx):=\sum_{(i,j)\in\ccalE} A_{ij} |x_i - x_j|$, which can be understood as a convex relation of the former. 

More involved approaches to relate the properties of the signals with the supporting graph consider that the signal $\bbx$ is bandlimited on a frequency domain given by the eigenvectors of the so-called graph shift operator. Such an operator is an $N\times N$ matrix typically set to either the adjacency $\bbA$ or the Laplacian $\bbL$ of $\ccalG$. Regardless of the particular choice, suppose that the selected matrix is diagonalizable and let $\bbV=[\bbv_1,...,\bbv_N]$ collect its $N$ eigenvectors. 
Then, $\bbV^{-1}$ can be interpreted as the graph Fourier transform (GFT) for graph signals and $\bbx$ is said to be $K$-bandlimited if $\tbx = \bbV^{-1}\bbx$ satisfies that $\tilde{x}_k=0$ for all $k>K$. 
Assuming that the graph is undirected (so that its matrix representations are symmetric and we have that $\bbV^{-1}=\bbV^T$), let us define
\begin{equation}\label{E:bandlimited_laplacian_quad}
\mathrm{BL}_\ccalG^K(\bbx):=\big\|[\bbv_{K+1},...,\bbv_N]^T\bbx\big\|_2^2.
\end{equation}   
Clearly, $\mathrm{BL}_\ccalG^K(\bbx)$ measures the energy of $\bbx$ present in the frequencies above $K$ and we have that if $\bbx$ is purely $K$-bandlimited,  then $\mathrm{BL}_\ccalG^K(\bbx)=0$. 

\subsection{Problem statement}
Let $\bbx\in\reals^N$ be a graph signal defined on a graph $\ccalG$. Suppose that $\bbx$ is not observed at all the nodes but only at a subset of them. Let $\bbm\in \{0,1\}^N$ denote a mask vector such that $m_i=1$ if the signal is observed at node $i$ and $m_i=0$, otherwise. Furthermore, we consider that the available observations do not correspond to the actual values of the signal $\bbx$, but to $\bbs \in \{-1,1\}^N$ a one-bit quantized (signed) version of it defined as $s_i = \mathrm{sign}(x_i)$. The observed signal is then
\begin{equation}\label{E:masked_signed_measurements}
\bar{\bbs}=\bbm \odot \bbs = \bbm \odot \mathrm{sign}(\bbx),
\end{equation}
where $\odot$ denotes the entry-wise product, $\mathrm{sign}$ is also applied entry-wise, and the entries of $\bar{\bbs}$ are either $-1$, $0$ or $1$. 
Our goal is to design an architecture that, using $\bar{\bbs}$ as input, is able to estimate $\bbx$. 
To do so, we assume that we are given a collection of $R$ observations $\{\bar{\bbs}^{(r)}\}_{r=1}^R$ each of them corresponding to an independent mask $\bbm^{(r)}$ and a signal $\bbx^{(r)}$ independently drawn from a common distribution that is either smooth or bandlimited in $\ccalG$ [cf. \eqref{E:smoothness_laplacian_quad} and \eqref{E:bandlimited_laplacian_quad}]. With this notation in place, we can formally state our problem.

\begin{Problem}
	Given a set of quantized graph signals with missing values $\{\bar{\bbs}^{(r)}\}_{r=1}^R$ generated as in \eqref{E:masked_signed_measurements}, estimate the complete and continuous associated signals $\{{\bbx}^{(r)}\}_{r=1}^R$.
\end{Problem}

It should be noted that the proposed problem is more challenging than what it appears to be at first sight.
First, even though the true signals ${\bbx}^{(r)}$ are assumed to be shaped by the graph, the (non-linear) quantization precludes a straightforward application of the associated graph regularizers. 
Second, and more importantly, \emph{at no point we have access to any complete and continuous signal} ${\bbx}^{(r)}$, potentially compromising the success of learning-based methods due to the access to a relatively poor training set. 
As will be presented next, a GAN-based approach can achieve satisfactory results even in this unfavorable scenario.

\section{Generative adversarial network for graph signal imputation}
\label{sec:GAMDIN}

\begin{figure}[t]
	\centering
	\includegraphics[width=\linewidth]{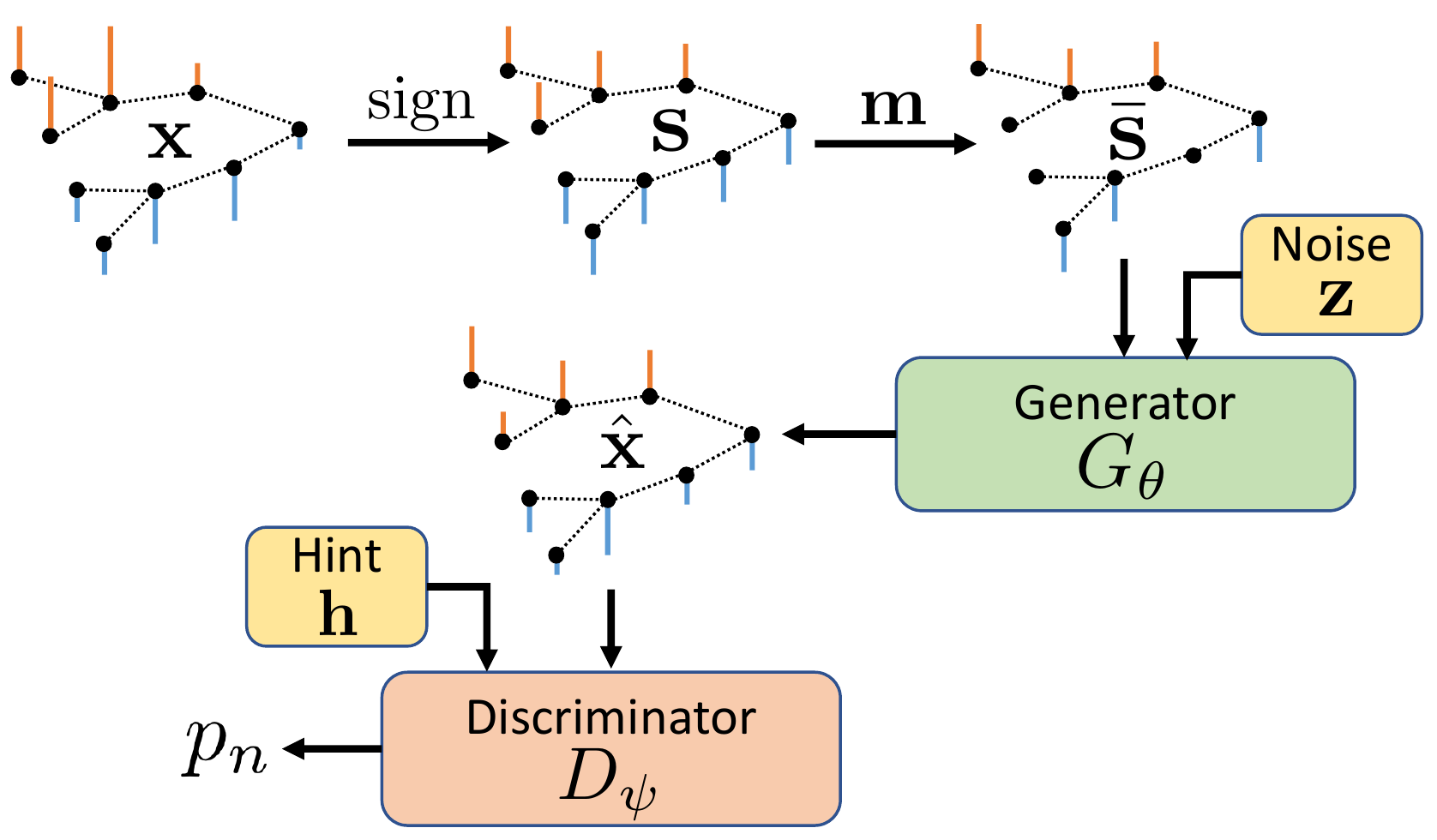}
	\caption{A schematic view of the proposed GAN architecture for a single graph signal $\bbx$. The signal is first quantized and then masked to obtain the observed signal $\bar{\bbs}$ [cf.~\eqref{E:masked_signed_measurements}]. The generator $G_\theta$ processes $\bar{\bbs}$ along with random noise $\bbz$ to obtain an estimate $\hat{\bbx}$ of the original signal $\bbx$. The discriminator $D_\psi$ is given a signed version of $\hat{\bbx}$ along with a hint $\bbh$ containing the true masking information for every node except for a randomly chosen node $n$. Based on this information, $D_\psi$ seeks to estimate the probability that node $n$ was indeed observed by the generator.
	}\label{fig:architecture}
\end{figure}

We now introduce our GAN approach for signal imputation inspired by~\cite{yoon2018gain}.
The main differences with this prior work lie on the consideration of graph regularizers for the generator and the quantization model for the masked signal. 
As will be seen in Section~\ref{results}, the introduced modifications improve the performance for the considered cases. 
Our architecture is summarized in Fig.~\ref{fig:architecture}.

The generator takes as inputs the observed realizations $\bar{\bbs} \in \{-1,0,1\}^N$ (we drop the superscript $^{(r)}$ in this section for conciseness) as well as independent white Gaussian noise $\bbz \in \reals^N$ that will be used to drive the imputation of the missing entries in $\bar{\bbs}$ (those with value zero).
Notice that the mask $\bbm$ can be inferred from $\bar{\bbs}$, i.e., $m_i = 1$ if $|s_i| = 1$ and $m_i = 0$ otherwise.
The output of the generator function $G$ is then given by
\begin{equation}\label{E:generator}
\hat{\bbx} = G_\theta({\bar{\bbs} }, (\mathbf{1}-\bbm) \odot \bbz),
\end{equation}
which seeks to replicate the true underlying signal $\bbx$ [cf.~\eqref{E:masked_signed_measurements}].
In this paper, we represent the generator function as a feedforward neural network $G_\theta$ with trainable parameters that are set via backpropagation to minimize the loss that will be introduced in~\eqref{E:generator_loss_1}-\eqref{E:generator_loss_3}.
However, since the loss of the generator depends on the performance of the discriminator, let us introduce the discriminator first.

\begin{figure*}[!th]
\centering
        \begin{subfigure}[b]{0.24\textwidth}
        
                \includegraphics[width=\linewidth]{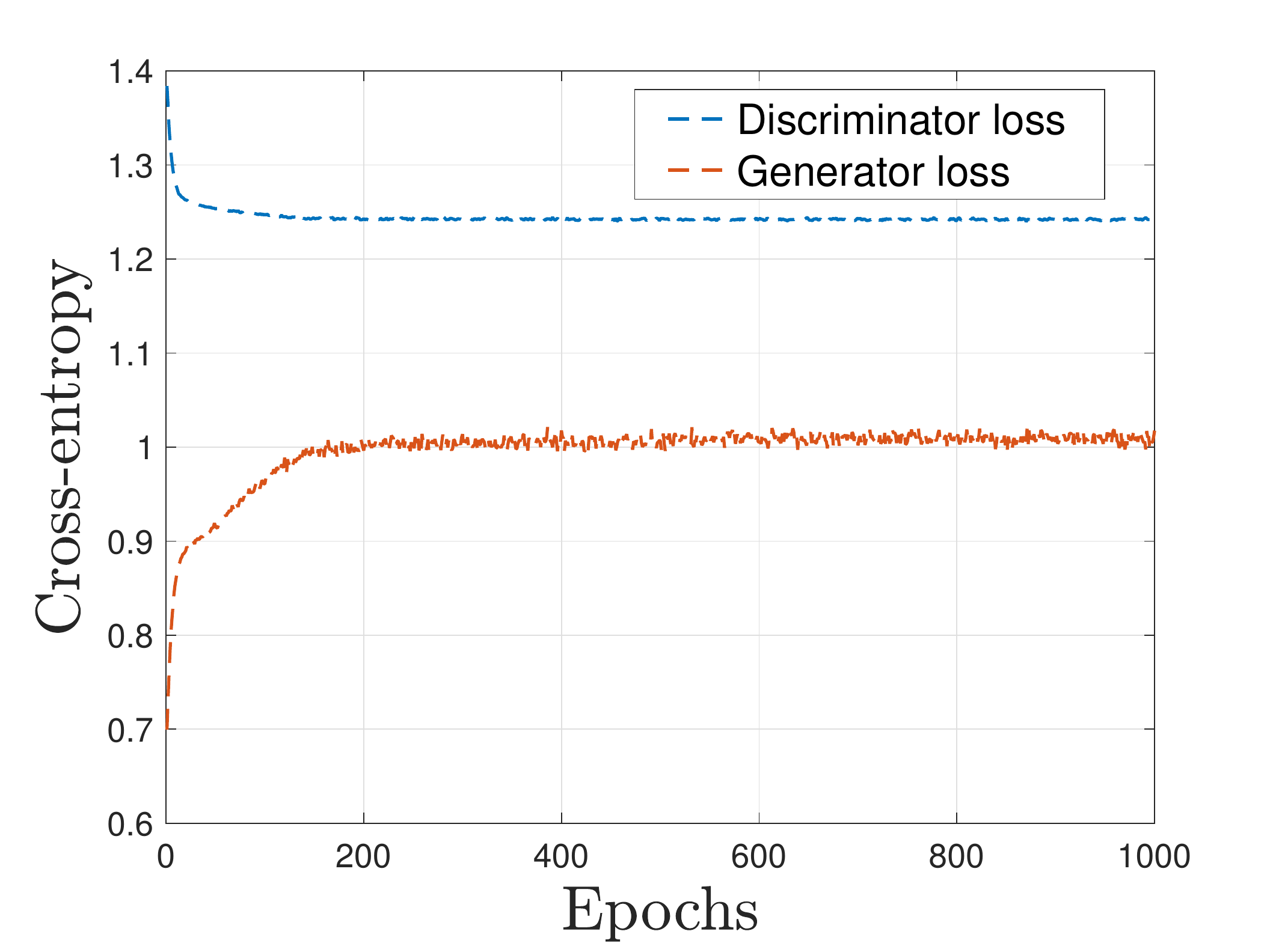}
                \caption{Proposed method}
                \label{fig:Cross_entropy_proposed}
        \end{subfigure}%
~     
        \begin{subfigure}[b]{0.24\textwidth}
        
                \includegraphics[width=\linewidth]{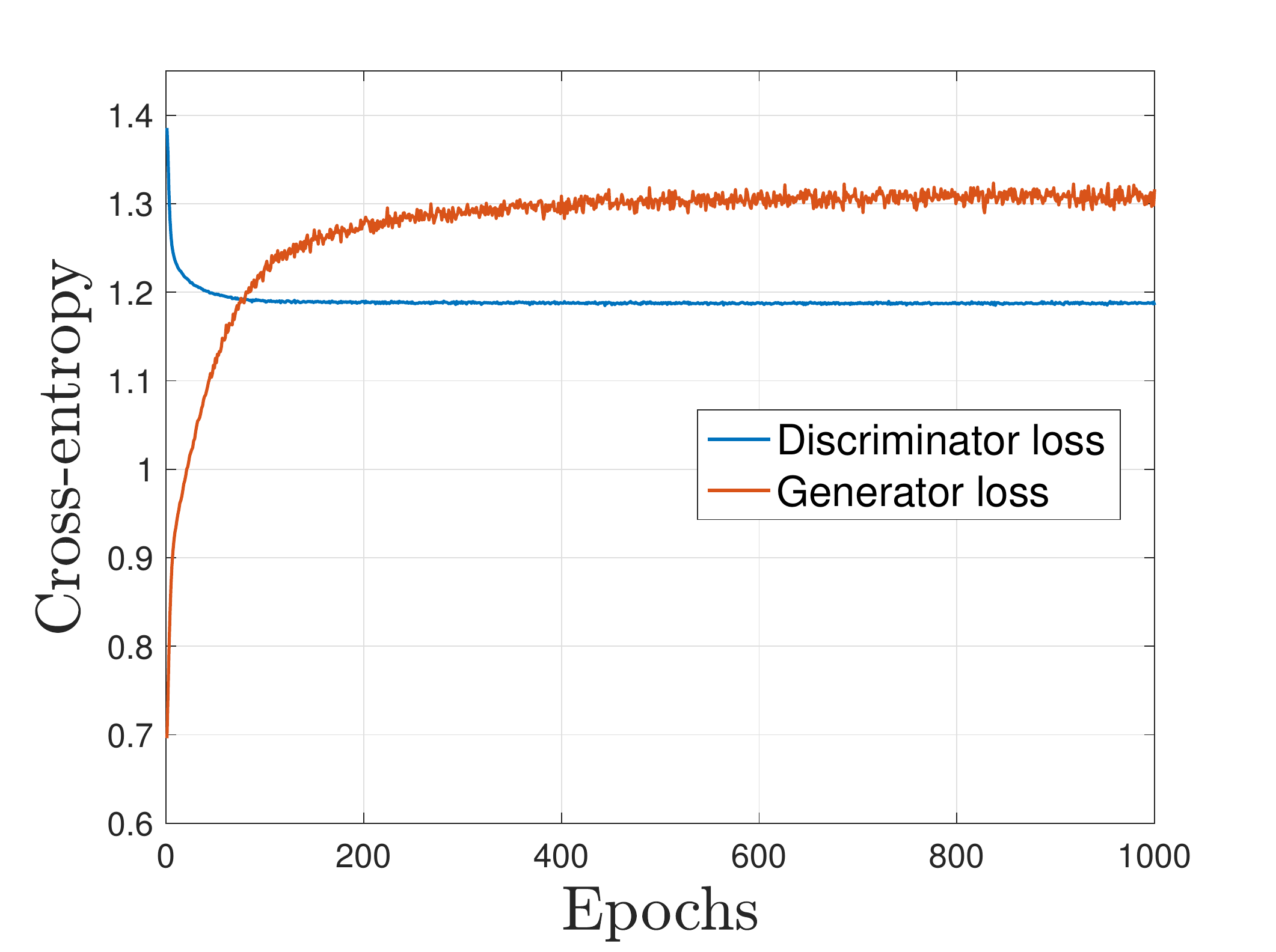}
                \caption{GAIN}
                \label{fig:Cross_entropy_Mihaela}
        \end{subfigure}  
~
       \begin{subfigure}[b]{0.24\textwidth}
                \includegraphics[width=\linewidth]{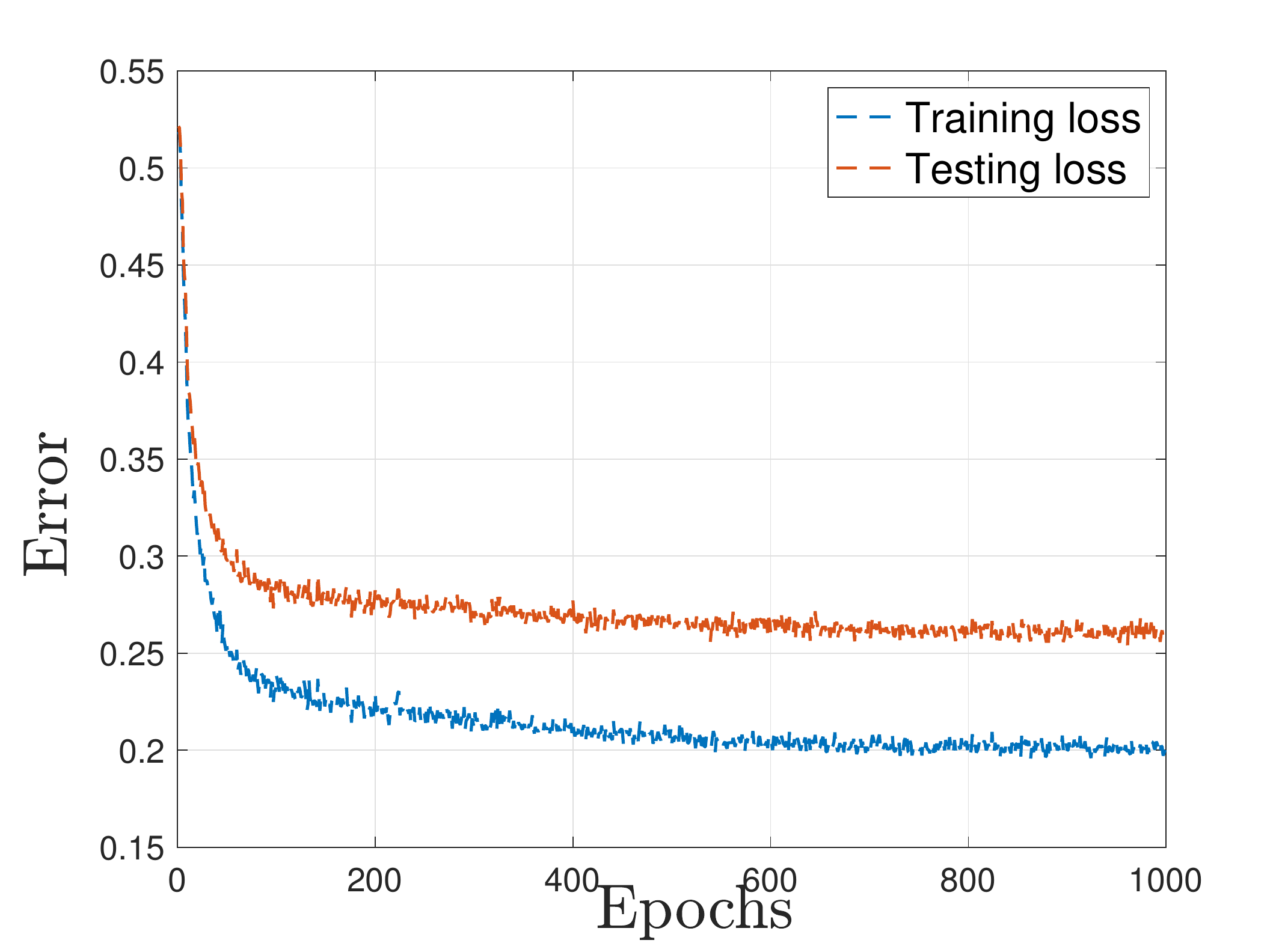}
                \caption{Proposed method}
                \label{fig:RMSE_measured_proposed}
        \end{subfigure}%
~        
        \begin{subfigure}[b]{0.24\textwidth}
                \includegraphics[width=\linewidth]{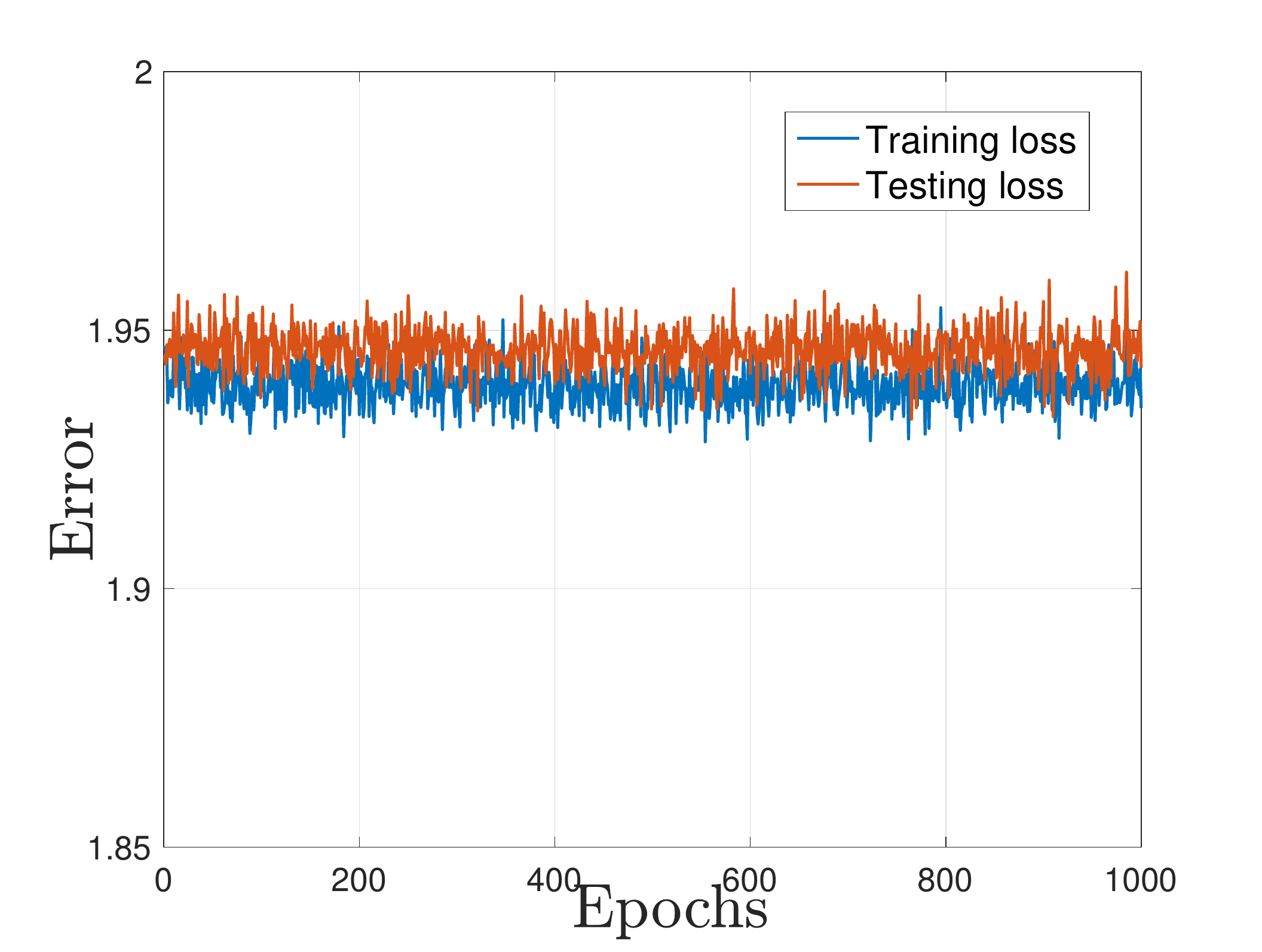}
                \caption{GAIN}
                \label{fig:RMSE_measured_Mihaela}
        \end{subfigure}
        \caption{Learning curves. (a) and (b) Cross-entropy loss of the generator and discriminator networks. The results in (a) and (b) demonstrate that the discriminator and generator losses in the proposed method converge faster than those in GAIN. (c) and (d) RMSE training and test loss. The results from (c) and (d) show that the proposed GAN architecture learns and imputes the missing data accurately, outperforming~GAIN.}   
 \label{fig:learning_curves}    
\end{figure*}

The objective of the discriminator is to detect the nodes with missing entries in the original graph signal that served as input to the generator, i.e., it tries to detect the positions of the zeros in $\bar{\bbs}$ from $\mathrm{sign}(\hat{\bbx})$. 
Intuitively, if the discriminator cannot identify these nodes, it implies that the generator has imputed the data correctly, since it is hard to tell apart the true observations from the imputations.
The task at the discriminator is especially hard given that at no point we have fully observed training data.
Hence, we follow the idea of providing a hint to the discriminator advocated in~\cite{yoon2018gain}.
More precisely, for each signal we choose one node uniformly at random, say node $n$, and then generate the hint vector $\bbh \in \{0, 0.5, 1\}^N$ where $h_n = 0.5$ and $h_{n'} = m_{n'}$ for all ${n'} \neq n$. 
Given $\bbh$, the discriminator knows the true mask values for every node except for node $n$ and needs to determine if the original signal was observed at node $n$ or not.
Formally, the discriminator function $D$ is given by
\begin{equation}\label{E:discriminator}
p_n = D_\psi(\mathrm{sign}(\hat{\bbx}), \bbh),
\end{equation}
where $p_n$ is the assigned probability that the value at node $n$ was indeed observed by the generator, i.e., the probability that $m_n = 1$. 
As done for the generator, we represent the discriminator as a feedforward neural network with parameters~$\psi$. 
As a loss function for the discriminator we consider the well-established cross entropy
\begin{equation}\label{E:cross_entropy}
\ccalL^D_{\theta,\psi}(p_n, m_n) = - [m_n \log(p_n) + (1-m_n) \log(1-p_n)],
\end{equation}
where we have made explicit the dependence on the neural network parameters $\theta$ and $\psi$ of both the generator and the discriminator.

Regarding the loss function of the generator, this will consist of three terms
\begin{align}
\ccalL^{G_1}_{\theta,\psi}(p_n,m_n) &= - (1-m_n) \log(p_n), \label{E:generator_loss_1} \\
\ccalL^{G_2}_{\theta}(\bar{\bbs}, \hat{\bbx}) &= \sum_{i=1}^N m_i (\bar{s}_i - \mathrm{sign}(\hat{x}_i))^2, \label{E:generator_loss_2} \\
\ccalL^{G_3}_{\theta}(\hat{\bbx}) &= \mathrm{TV}_\ccalG^{\ell_2}(\hat{\bbx}). \label{E:generator_loss_3}
\end{align}
The loss in \eqref{E:generator_loss_1} is minimized when the generator is able to deceive the discriminator, i.e., when this latter assigns a high probability $p_n$ to a node that was not observed $m_n=0$. 
The loss in \eqref{E:generator_loss_2} promotes that the signs of the output of the generator $\hat{\bbx}$ coincide with the true observed signs in $\bar{\bbs}$. 
Finally, the loss in \eqref{E:generator_loss_3} enforces that the generated signal $\hat{\bbx}$ is smooth on the underlying graph [cf.~\eqref{E:smoothness_laplacian_quad}].
As discussed in Section~\ref{sec:problem}, this loss can be replaced by other graph regularizers in order to promote different signal models.
Notice that the losses in \eqref{E:generator_loss_2} and \eqref{E:generator_loss_3} do not depend on the discriminator and simply force the generator to create smooth signals that have the right sign pattern. By contrast, the loss in \eqref{E:generator_loss_1} ties the generator and the discriminator performances giving rise to the classical adversarial  setting of GANs. More precisely, we define our objective in terms of the minimax problem
\begin{align}\label{E:minimax}
\min_\theta \max_\psi \,\,\, &-\ccalL^D_{\theta,\psi}(p_n, m_n) + \ccalL^{G_1}_{\theta,\psi}(p_n,m_n) \nonumber \\
& + \alpha \ccalL^{G_2}_{\theta}(\bar{\bbs}, \hat{\bbx}) + \beta \ccalL^{G_3}_{\theta}(\hat{\bbx}),
\end{align}
where the scalars $\alpha$ and $\beta$ control the trade-off between the different components of the loss function at the generator.
As typically done, we optimize \eqref{E:minimax} in an iterative manner where we first fix $\theta$ and update the parameters of the discriminator via mini-batch training, followed by fixing $\psi$ and updating the values of $\theta$ accordingly.
More details on the implementation and hyperparameter selection are given in the next section.

\section{Numerical Results}
\label{results}

The generator and discriminator networks are modeled as fully connected neural networks each having a depth of 3 layers (including the output layer). 
The number of hidden nodes in each layer is $256$, $128$, and $784$, respectively.
We use ${\tt tanh(\cdot)}$ as the nonlinear activation function at all layers of the generator network. 
For the discriminator network, we use ${\tt tanh(\cdot)}$ as the activation function at all the intermediate layers, while we use the ${\tt sigmoid}(\cdot)$ activation function at the output layer. 
The hyperparameters $\alpha$ and $\beta$ in \eqref{E:minimax} are selected via cross-validation. 
The proposed architecture is implemented in Python using the TensorFlow API~\cite{abadi2016tensorflow} with the ADAM stochastic optimizer~\cite{kingma2014adam}. 

\begin{figure*}[!th]
\centering
        \begin{subfigure}[b]{0.33\textwidth}
                \includegraphics[width=\linewidth]{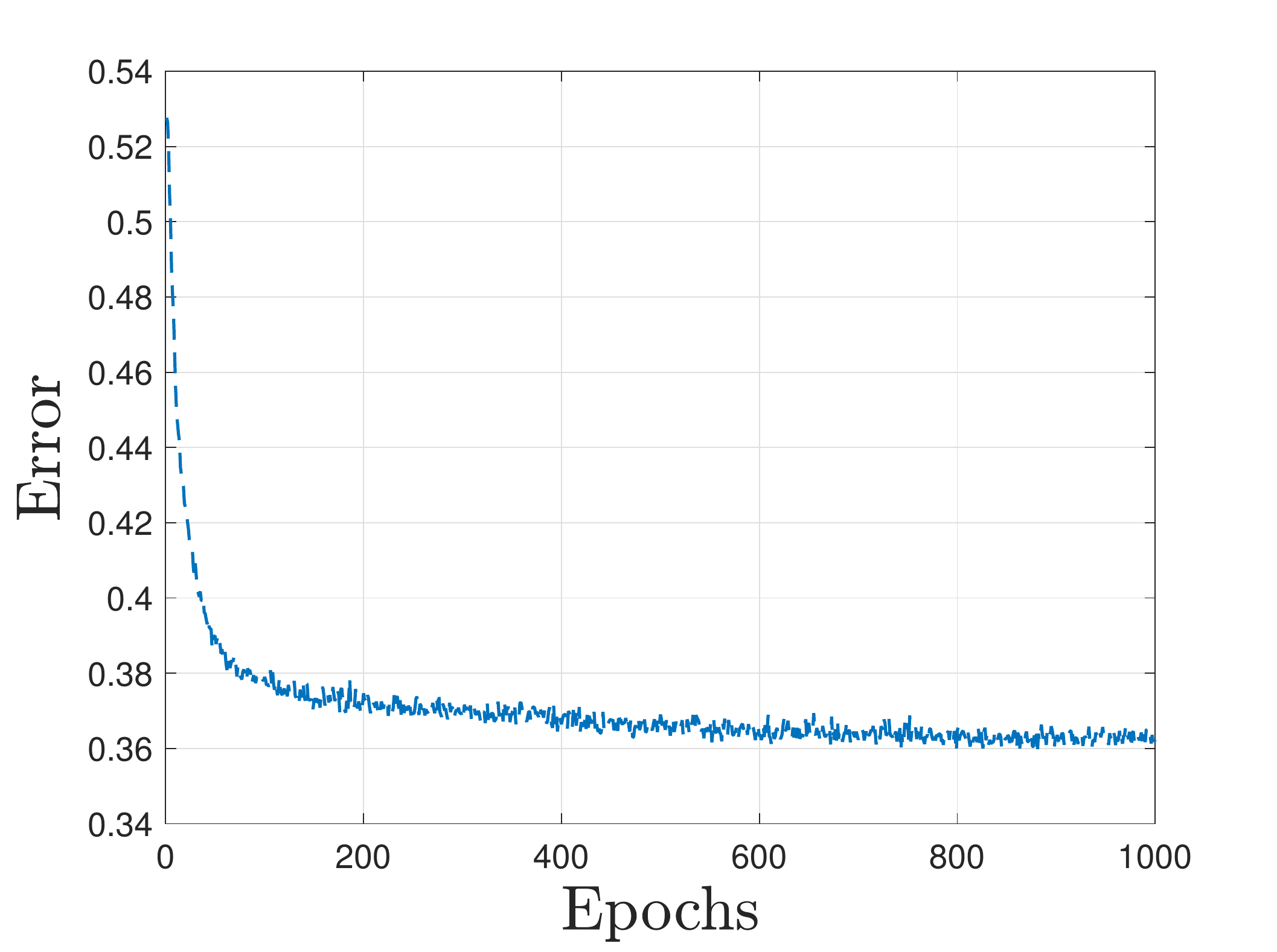}
                \caption{Proposed method}
                \label{RMSE_proposed}
        \end{subfigure}%
        \begin{subfigure}[b]{0.33\textwidth}
 ~       
                \includegraphics[width=\linewidth]{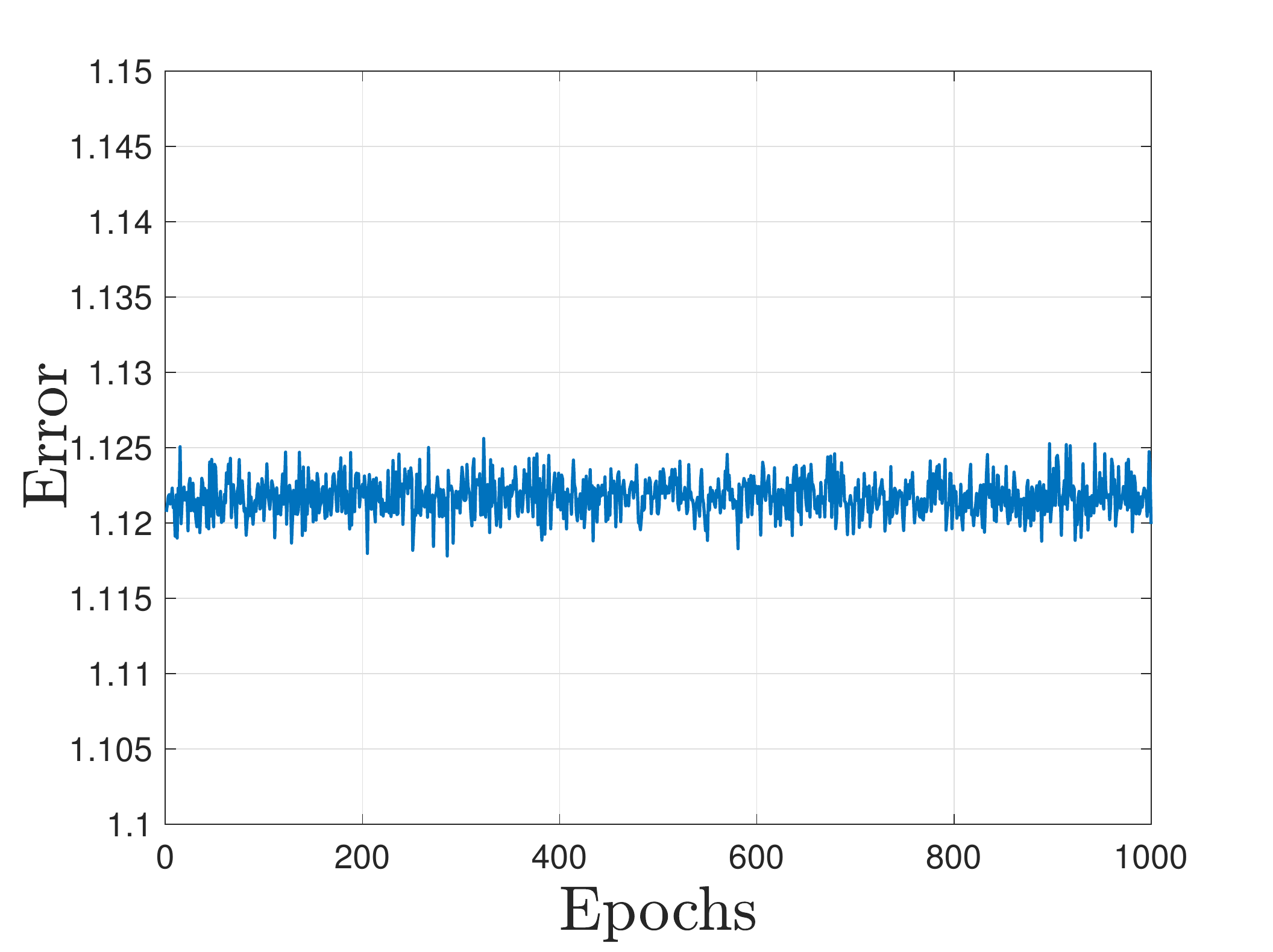}
                \caption{GAIN}
                \label{RMSE_Mihaela}
        \end{subfigure}      
        \begin{subfigure}[b]{0.33\textwidth}
~        
                \includegraphics[width=\linewidth]{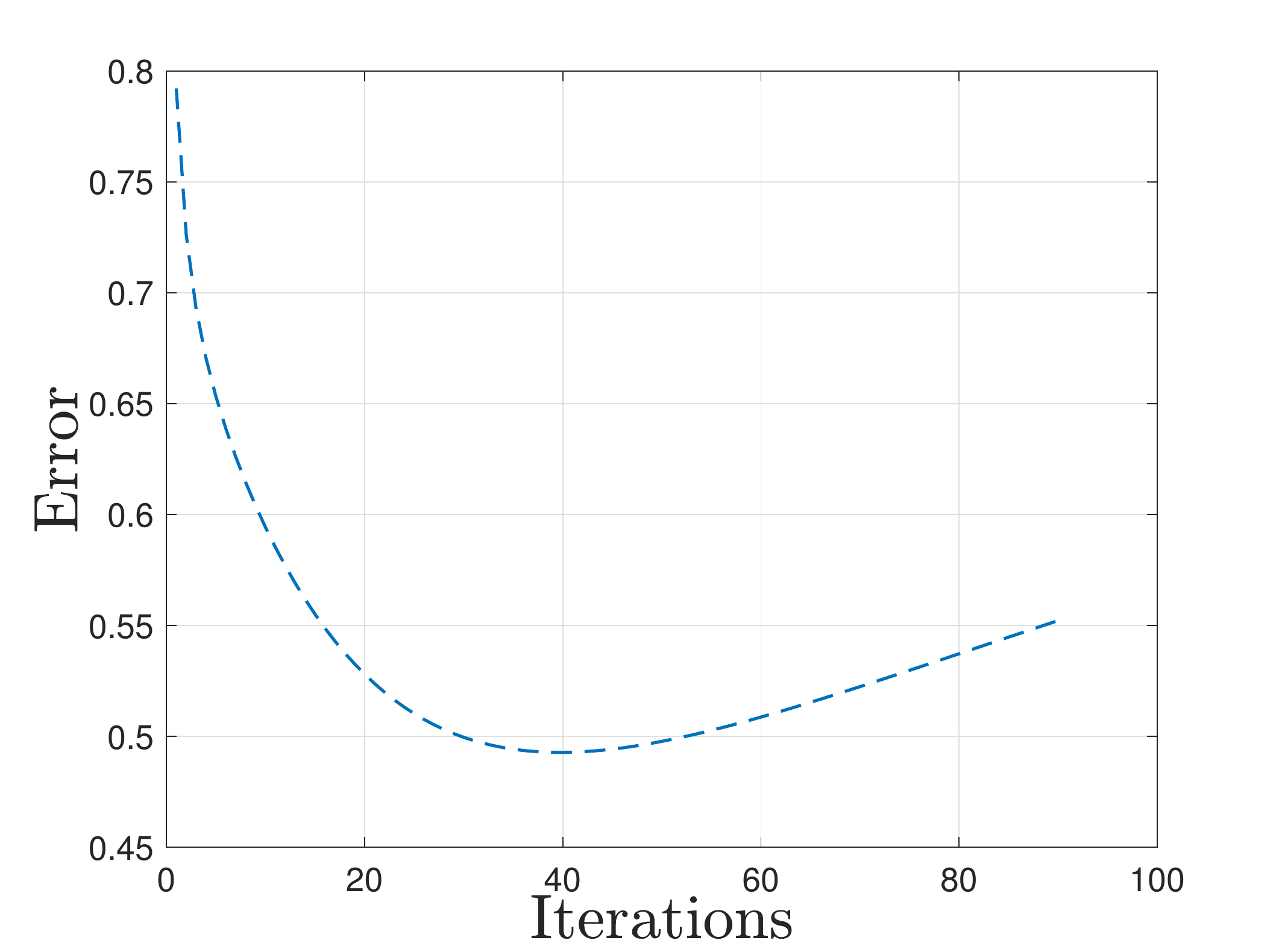}
                \caption{Gradient descent}
                \label{RMSE_non_deep}
        \end{subfigure}
        \caption{Error between the imputed and the ground truth signals for the three compared methods. 
        The decrease in error observed in (a) is a manifestation of effective learning for our proposed method.}  
        \label{RMSE}
                	\vspace*{-5mm}
\end{figure*}

In this section, we evaluate the performance of the proposed method using a real-world dataset, namely, the MNIST handwritten digits dataset ~\cite{lecun2010mnist}. We consider $55,000$ images for training and $10,000$ images for testing. Each image of size $28\times 28$ pixels is vectorized to yield a vector $\bbx \in \mathbb{R}^{784}$. We use the $k$-nearest neighbors method with a nodal degree of $20$ to construct a graph with $784$ nodes (pixels) from the available training data. We consider data with $50\%$ missing entries for all the experiments. The entries of the mask are generated independently and identically (across both pixels and experiments) using a Bernoulli distribution with a success probability of 0.5.

To demonstrate the performance of the proposed method, we consider: i) GAIN \cite{yoon2018gain}, which is obtained by setting the tuning parameter $\beta=0$, and ii) a non-deep learning approach based on gradient descent.
Specifically, we use the following update equation for the gradient descent 
\[
\bbx^{(i+1)} = \bbx^{(i)} - \mu \nabla J(\bbx^{(i)}),
\] 
where $\mu$ is the step size. Here, $\nabla J(\bbx)$ is the gradient vector of the graph Laplacian regularized loss function $J(\bbx) = \|\bar{\bbs} - \bbm \odot \tanh(\bbx)\|_2^2 + \beta \mathrm{TV}_\ccalG^{\ell_2}(\bbx)$, where we use $\tanh(\cdot)$ to approximate $\sign(\cdot)$. We use $\mu = 0.01$ and stop the gradient descent update at the $40$th iteration after which the error increases; see Fig.~\ref{RMSE_non_deep}.

Fig.~\ref{fig:learning_curves} plots different learning curves. In Fig.~\ref{fig:Cross_entropy_proposed} the convergence of the generator and discriminator cross-entropies for the proposed architecture are shown. Since the generator loss is clearly below the discriminator loss, the discriminator fails to distinguish between the imputed and observed values. In the case of GAIN, although the loss of the discriminator converges, as can be seen in Fig.~\ref{fig:Cross_entropy_Mihaela}, the generator loss is larger than the discriminator loss, revealing that the generator fails to impute the missing entries accurately.  

Fig.~\ref{fig:RMSE_measured_proposed} and  Fig.~\ref{fig:RMSE_measured_Mihaela} show the behavior of the learning curves across epochs. 
The  mean-squared error (averaged over the training data) between the observed and imputed data at the pixels corresponding to the observed data is referred to as {\it training loss}. Further, the mean-squared error (averaged over the training data) between the imputed data and the ground truth at the missing entries 
is referred to as {\it testing loss}. Fig.~\ref{fig:RMSE_measured_proposed} shows that the mean-squared error decreases as the number of epochs increases, which indicates that the generator is learning the underlying data distribution and imputing the missing entries accurately. In the case of GAIN, both the training and testing losses are not improving with the number of epochs. This means that GAIN is not learning the underlying data distribution. 
 
\begin{figure}[!t]
\centering
	\includegraphics[width=\columnwidth]{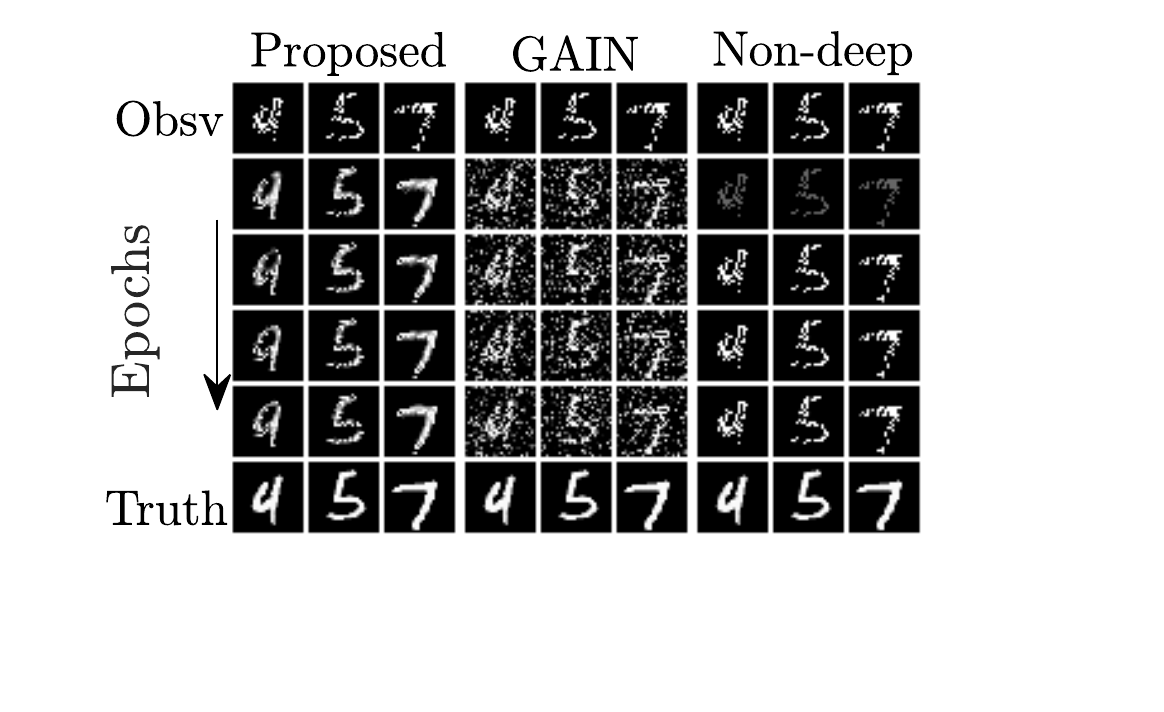}
	        	\vspace*{-8mm}
        \caption{Recovery of MNIST digits from data with $50\%$ missing entries. The first row represents the observed data with $50\%$ missing entries. The second to fifth rows of the image represent the recovered images for different epochs ($250,500,750,1000$) or iterations. The last row represents the ground truth. The proposed method imputes the missing data accurately as compared to the other approaches.}
        \label{mnist_data}
\end{figure}

Fig.~\ref{RMSE} illustrates the mean-squared error (averaged over 10000 test images) between the recovered data and the ground truth for different epochs, while for Fig.~\ref{RMSE_non_deep} to compute the mean-squared error we averaged over 100 test images. Fig.~\ref{RMSE_proposed} shows that the error converges for the proposed method. Whereas, in the case of GAIN and the gradient descent method, the error  either does not converge or converges to a very high level, as shown in Fig.~\ref{RMSE_Mihaela} and Fig.~\ref{RMSE_non_deep}, respectively.
 

The results of the imputed pixels along with observed data are shown in Fig.~\ref{mnist_data}. 
The first row shows the observed data with $50\%$ missing entries. The rows from second to fifth show the recovered images for a different number of epochs ($250,500,750,1000$) or iterations ($10, 20, 30, 50$) in the case of the gradient descent method. The last row represents the ground truth. As the number of epochs increases, the proposed method imputes the missing data more accurately as compared to GAIN and the iterative gradient descent method.  
 
Table~\ref{rmse_avg}, shows the average error between the imputed test images and the ground truth. One can observe that the average error of the proposed method is significantly smaller than that of the considered baseline approaches. 
 \begin{table}[!t]
\centering
\caption{Error comparison averaged across $10,000$ test samples with $50\%$ missing entries.}
\label{rmse_avg}
\begin{tabular}{|c|c|}
\hline 
Method   &   Error   \\ \hline
    \textbf{Proposed}        &  $\textbf{0.36}$                             \\ \hline
   GAIN  \cite{yoon2018gain}      &       $1.12$                          \\ \hline
 Iterative gradient descent           &       $ 0.49$                               \\ \hline
\end{tabular}
\vspace*{-4mm}
\end{table}

\section{Conclusions and future work}
\label{sec:majhead}

We developed a deep learning method for the recovery of smooth graph signals from quantized observations and missing data. 
The success of the proposed method hinges on combining the flexibility and proven empirical value of GANs with graph regularizers that promote the desired signal structure.
Current and future work includes consideration of multi-bit quantization and non-scalar observation models, exhaustive comparisons with other reconstruction methods, and consideration of architectures that incorporate the underlying graph structure in the functional forms of the generator and the discriminator, thus departing from feedforward neural networks and relying on more specialized graph convolutional networks.

\vfill\pagebreak


\bibliographystyle{IEEEbib}
\bibliography{strings,refs}

\end{document}